\documentclass[aps,prb,twocolumn,showpacs,preprintnumbers,amsmath,amssymb,superscriptaddress]{revtex4}%

\usepackage{graphicx}%
\usepackage{dcolumn}
\usepackage{amsmath}

\makeatletter
\def\btt#1{\texttt{\@backslashchar#1}}%
\DeclareRobustCommand\bblash{\btt{\@backslashchar}}%
\makeatother

\begin{document}

\preprint{PREPRINT (\today)}

\title{Pressure effects on the superconducting properties
of YBa$_{2}$Cu$_{4}$O$_{8}$}

\author{R.~Khasanov}
\affiliation{ Laboratory for Neutron Scattering, ETH Z\"urich and
Paul Scherrer Institut, CH-5232 Villigen PSI, Switzerland}
\affiliation{DPMC, Universit\'e de Gen\`eve, 24 Quai
Ernest-Ansermet, 1211 Gen\`eve 4, Switzerland}
\affiliation{Physik-Institut der Universit\"{a}t Z\"{u}rich,
Winterthurerstrasse 190, CH-8057, Z\"urich, Switzerland}
\author{T.~Schneider}
\affiliation{Physik-Institut der Universit\"{a}t Z\"{u}rich,
Winterthurerstrasse 190, CH-8057, Z\"urich, Switzerland}

\author{H.~Keller}
\affiliation{Physik-Institut der Universit\"{a}t Z\"{u}rich,
Winterthurerstrasse 190, CH-8057, Z\"urich, Switzerland}

\begin{abstract}
Measurements of the magnetization under high hydrostatic pressure
(up to 10.2~kbar) in  YBa$_{2}$Cu$_{4}$O$_{8}$ were carried out.
%
%
From the scaling analysis of the magnetization data  the pressure
induced shifts of the transition temperature $T_{c}$, the volume
$V$ and the anisotropy $\gamma$ have been obtained.
It was shown that the pressure induced relative shift of $T_{c}$
mirrors essentially that of the anisotropy. This observation
uncovers a novel generic property of anisotropic type II
superconductors, that inexistent in the isotropic case.

\end{abstract}

\maketitle

In the cuprate high-temperature superconductors (HTS), the
canonical change in $T_{c}$ with pressure is that it first
increases, passes through a maximum value at some critical
pressure, and then decreases.\cite{takahashi,wijngaarden} It has
been argued that there are at least two effects determining the
total pressure dependence of $T_{c}$, an intrinsic effect and the
one that arises from the pressure-induced changes in the charge
carrier concentration.\cite {takahashi,wijngaarden} However, there
are effects which make the discrimination between these
contributions difficult. In most HTS the application of pressure
doesn't simply compress the lattice, but also prompts mobile
oxygen defects to assume a greater degree of local
order.\cite{klehe} This leads to relaxation effects which are both
temperature and pressure dependent. An exception is the
double-chain compound YBa$_{2}$Cu$_{4}$O$_{8}$  with a fixed
oxygen stoichiometry. Bucher {\it et al.} [\onlinecite{bucher}]
found that in this compound $T_{c}$ increases under hydrostatic
pressure with the rate ${\rm d}T_{c}/{\rm d}p\simeq 5.5$~K/GPa.
Subsequent studies revealed that with increasing pressure $T_{c}$
passes through a maximum around 9~GPa and then
decreases.\cite{takahashi,wijngaarden,klehe} Furthermore,
measurements of the in-plane penetration depth $\lambda_{ab}$ of
YBa$_{2}$Cu$_{4}$O$_{8}$ revealed pressure induced changes which
cannot be simply attributed to the pressure changes of
$T_{c}$.\cite{khasanov,khasanov2} Since in cuprate
high-temperature superconductors, including
YBa$_{2}$Cu$_{4}$O$_{8}$, the critical regime where thermal
critical fluctuations dominate is experimentally accessible,
various critical properties are not independent but related by
universal
realtions.\cite{tshk,tsjh,book,parks,ts03,bled,tshknj,tspstat}
Accordingly, the isotope or pressure effects on these quantities
are related as well.\cite{ts03,tspstat} Here we explore the
universal relationship between the pressure effects on transition
temperature $T_c$, volume $V$ and anisotropy $\gamma$ emerging
from the pressure induced changes of the magnetization near
$T_{c}$. It is shown that in the underdoped
YBa$_{2}$Cu$_{4}$O$_{8}$ the pressure effect on $T_{c}$ mirrors
essentially that on the anisotropy. This uncovers a novel generic
property of anisotropic type II superconductors, inexistent in the
isotropic case.
%

\begin{figure}[htb]
\includegraphics[width=1.05\linewidth]{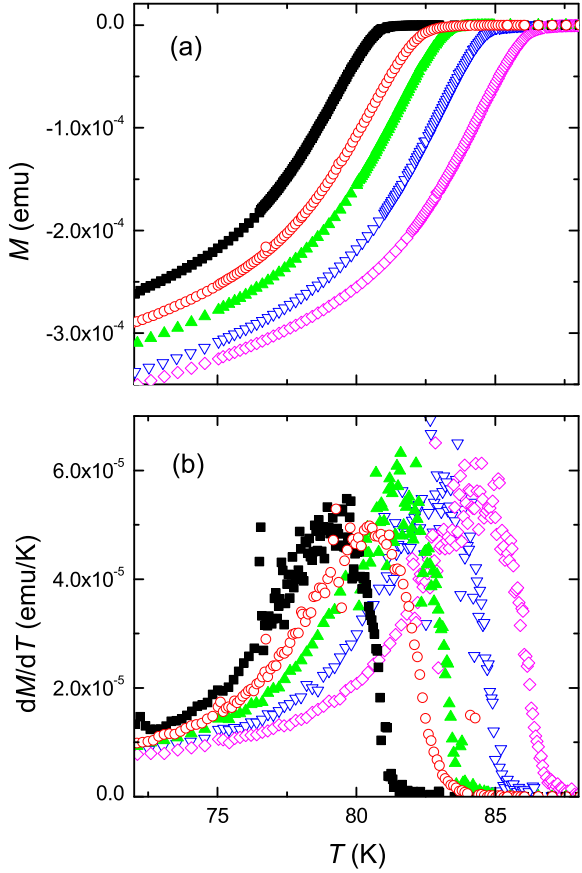}
 \caption{ (a) Field-cooled ($0.5$mT) magnetization
of a YBa$_{2}$Cu$_{4}$O$_{8}$ powder sample {\it vs}. $T$ near
$T_{c}$ for various applied hydrostatic pressures (from the left
to the right) 0.0~kbar, 2.67~kbar, 4.29~kbar, 7.52~kbar, and
10.2~kbar. (b) $dM/dT$ {\it vs}. $T$ for the data shown in
Fig.~\ref{fig1}(a) }
 \label{fig1}
\end{figure}

The polycrystalline YBa$_{2}$Cu$_{4}$O$_{8}$ samples were
synthesized by solid-state reactions using high-purity
Y$_{2}$O$_{3}$, BaCO$_{3}$ and CuO.\cite{khasanov} The hydrostatic
pressure was generated in a copper-beryllium piston cylinder clamp
that was especially designed for magnetization measurements under
pressure.\cite{strassle} The sample was thoroughly mixed with
Fluorient FC77 (pressure transmitting medium) with a sample to
liquid volume ratio of approximately $1/6$. The pressure was
measured in situ by monitoring the $T_{c}$ shift of the small
piece of In [$T_{c}(p=0)\simeq 3.4K$] included in the pressure
cell. The field-cooled magnetization  measurements were performed
with a Quantum Design SQUID magnetometer at temperatures ranging
from $2$ to $100$K.
In Fig.~\ref{fig1}(a) we displayed the field-cooled ($0.5$~mT)
magnetization $M$ of a YBa$_{2}$Cu$_{4}$O$_{8}$ powder sample {\it
vs}. $T$ near $T_{c}$ for various applied hydrostatic pressures
(0.0~kbar, 2.67~kbar, 4.29~kbar, 7.52~kbar, and 10.2~kbar). To
identify the temperature regime where critical fluctuations play
an essential role it is instructive to consider the behavior of
${\rm d}M/{\rm d}T$ \textit{vs.} $T$ displayed in
Fig.~\ref{fig1}(b). With decreasing temperature ${\rm d}M/{\rm
d}T$ is seen to raise below the transition temperature and after
passing a maximum value it decreases. This behavior contradicts
the mean-field behavior where ${\rm d}M/{\rm d}T$ below $T_c$
adopts a constant value. \cite{abrikosov} It implies that the
fluctuation dominated regime is accessible and attained. This
calls for an analysis beyond the mean-field approximation as
outlined below.

The detailed description of the scaling analysis adopted for
magnetization changes caused by isotope substitution or applying
pressure can be found in Refs.~\onlinecite{Schneider04} and
\onlinecite{Schneider05}. Briefly, the basic principles of the
analysis can be summarized as follows. When three-dimensional (3D)
Gaussian or 3D-XY thermal fluctuations dominate, the combination
$m\left( T,\delta ,H\right) /\left( \gamma \epsilon ^{3/2}\left(
\delta \right) T\sqrt{H}\right) $ adopts at $T_{c}$ a fixed
value\cite{tsjh,book,parks,bled}
\begin{equation}
\frac{m\left( T_{c}\right) }{T_{c}\left[ \gamma \epsilon
^{3/2}\left( \gamma ,\delta \right) \right]
_{T_{c}}\sqrt{H}}=-\frac{k_{B}C}{\Phi _{0}^{3/2}}, \label{eq1}
\end{equation}
where $m=M/V$ is the magnetization per unit volume, $C$ a constant
adopting for Gaussian and 3D-XY fluctuations distinct universal
values. Furthermore, $\epsilon \left( \delta \right) =\left( \cos
^{2}\left( \delta \right) +\sin ^{2}\left( \delta \right) /\gamma
^{2}\right) ^{1/2}$, where $\delta $ is the angle between the
applied magnetic field $H$ and the $c$-axis, $\Phi _{0}$ the flux
quantum, and $k_{B}$  Boltzmann's constant.
In powder samples this relation reduces to
\begin{equation}
\frac{m\left( T_{c}\right) }{T_{c}\sqrt{H}f\left( \gamma \left(
T_{c}\right) \right) }=-\frac{k_{B}C}{\Phi _{0}^{3/2}},\text{
}f\left( \gamma \left( T_{c}\right) \right) =\left[ \gamma
\left\langle \epsilon ^{3/2}\left( \gamma ,\delta \right)
\right\rangle \right] _{T_{c}}.  \label{eq2}
\end{equation}
As the pressure effect on the magnetization at fixed magnetic
field is concerned it implies that the relative shifts of the
magnetization $M$, volume $V$, magnetization per unit volume $m$,
anisotropy $\gamma $ and $T_{c}$ are not independent but close to
$T_{c}$ related by
\begin{equation}
\frac{\Delta M}{M}=\frac{\Delta V}{V}+\frac{\Delta
T_{c}}{T_{c}}+\frac{\Delta f}{f}.  \label{eq3}
\end{equation}
For YBa$_{2}$Cu$_{4}$O$_{8}$, where $\gamma >>1$ (see {\it e.g.}
Ref.~\onlinecite{kagawa}) it reduces to
\begin{equation}
\frac{\Delta M}{M}=\frac{\Delta V}{V}+\frac{\Delta
T_{c}}{T_{c}}+\frac{\Delta \gamma }{\gamma }  \label{eq4}
\end{equation}
On that condition it is impossible to extract these changes from
the temperature dependence of the magnetization. However,
supposing that close to criticality the magnetization data scale
within experimental error as
\begin{equation}
^{p=0}M\left( T\right) =\text{ }^{p}M\left( aT\right) ,
\label{eq5}
\end{equation}
the universal relation (\ref{eq4}) reduces to
\begin{equation}
\frac{\Delta T_{c}}{T_{c}}=-\frac{\Delta V}{V}-\frac{\Delta \gamma
}{\gamma }=\frac{T_{c}\left( p\right) }{T_{c}\left( 0\right)
}-1=\frac{1}{a}-1. \label{eq6}
\end{equation}

\begin{figure}[htb]
\includegraphics[width=1.05\linewidth]{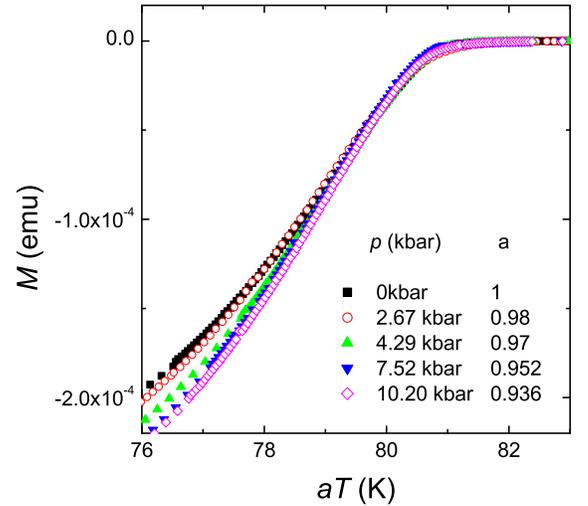}
 \caption{ Magnetization data rescaled according to $^{p=0}M\left(
T\right) =$ $^{p}M\left( aT\right) $ [Eq.(\ref{eq5})].  }
 \label{fig2}
\end{figure}

In Fig.~\ref{fig2} we displayed the magnetization data rescaled
according to Eq.(\ref{eq5}). Near $T_{c}$ the pressure induced
changes in $M(aT)$ is negligibly small so that Eqs.~(\ref{eq5})
and (\ref{eq6}) can be applied within experimental error.

\begin{table*}[htb]
\caption[~]{\label{Table:pressure_results} Estimates for the
pressure induced relative change of volume $ V$, transition
temperature $T_{c}$ and anisotropy $\gamma$ derived from the data
shown in Fig.~\ref{fig2} with the aid of Eqs.(\ref{eq5}), (\ref
{eq6}) and (\ref{eq7}). $T_{c}$ and $\lambda_{ab0}^{-2}$ at $p=0$,
4.29, 7.52, and 10.2~kbar taken from Khasanov \textit{et al}.
Ref.~\onlinecite{khasanov}. $T_{c}$ and $\lambda _{ab0}^{-2}$ at
$p=$1.2 and 2.67~kbar were obtained from the linear interpolation
of $T_c(p)$ and $\lambda_{ab0}^{-2}(p)$ data from
Ref.~\onlinecite{khasanov}. $\Delta \xi _{c0}/\xi _{c0}$ follows
from Eq.(\ref{eq9}) and $\Delta \xi _{ab0}/\xi _{ab0}$ from
relation (\ref{eq10}). Hereafter the relative pressure shift of
the physical quantity $X$ is determined as $\Delta
X(p)/X(0)=[X(p)-X(0)]/X(0)$.
} %
\begin{center}
\begin{tabular}{cccccccccccc}\\
 \hline
 \hline
 $p$ (kbar) & $\Delta V/V$ & $a$ & $\Delta T_{c}/T_{c}$ &
$\Delta \gamma /\gamma $ & $T_{c}$(K) & $\lambda _{ab0}^{-2}$($\mu
\text{m}^{-2}$) & $\Delta \xi _{co}/\xi _{c0}$ & $\Delta \xi
_{abo}/\xi _{ab0}$ &
$2\Delta \lambda_{abo}/\lambda _{ab0}$ \\
 \hline
0 & - & 1 & - & - & 79.07 & 34.3 & - & - & - \\
1.2&-0.0010&0.991&0.008&-0.007&79.67&35.9&-0.04&-0.05&-0.05 \\
2.67&-0.0023&0.98&0.017&-0.015&80.41&37.5&-0.07&-0.09&-0.09 \\
4.29&-0.0037&0.97&0.031&-0.027&81.4&39.2&-0.10&-0.13&-0.14 \\
7.52&-0.0064&0.952&0.050&-0.044&82.73&44.1&-0.19&-0.23&-0.28  \\
10.2&-0.0087&0.936&0.068&-0.059&84.22&45.0&-0.19&-0.25&-0.32  \\
 \hline \hline \\

\end{tabular}
   \end{center}
\end{table*}

\begin{figure}[tbp]
\centering
\includegraphics[width=1.05\linewidth]{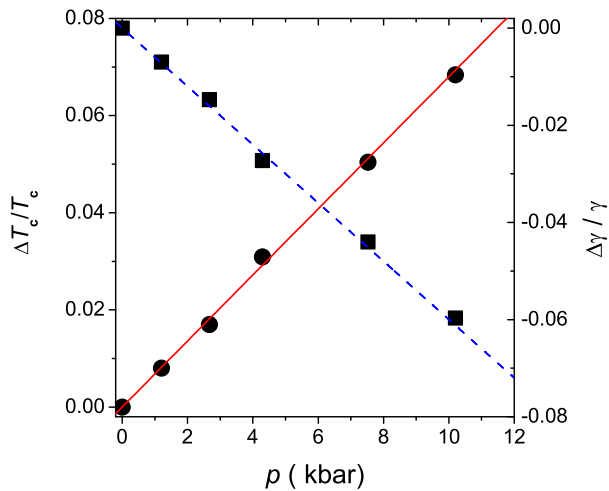}
\caption{$\Delta T_{c}/T_{c}$ ($\bullet $) and $\Delta \gamma
/\gamma $ ($\blacksquare $) {\it vs}. $p$ for the estimates listed
in Table~I.  The solid line is $\Delta T_{c}/T_{c}=0.0068p$ and
the dashed one is $\Delta \gamma /\gamma =-0.006p$ with $p$ in
kbar.}
 \label{fig3}
\end{figure}

Since $a$ is less than one and decreases with increasing pressure
the transition temperature increases in the pressure range
considered here. Given then the bulk modulus $B=1174$~kbar
(Ref.~\onlinecite{bucher}) for the pressure dependence of the
relative volume change we obtain the expression
\begin{equation}
\frac{\Delta V}{V}=-\frac{p\left( \text{kbar}\right) }{1174},
\label{eq7}
\end{equation}
and with Eq.(\ref{eq6}) the estimates for the pressure induced
changes of the anisotropy listed in
Table~\ref{Table:pressure_results}. It is readily seen that the
rise of $T_{c}$ mirrors essentially the decrease of the
anisotropy. Indeed the relative volume change is an order of
magnitude smaller. Furthermore, from Fig.~\ref{fig3}, showing
$\Delta T_{c}/T_{c}$ and $\Delta \gamma /\gamma $ versus $p$ for
the estimates listed in Table~\ref{Table:pressure_results}, it is
seen that in the pressure range considered here, both $\Delta
T_{c}/T_{c}$ and $\Delta \gamma /\gamma $ depend nearly linearly
on pressure.
Thus, under pressure not only $T_{c}$ and the volume changes but
the anisotropy is modified as well. A quick glance to
Table~\ref{Table:pressure_results} reveal that the pressure effect
on $T_{c}$ mirrors essentially that on the anisotropy.

In addition, recent in-plane penetration depth $\lambda_{ab}$
measurements revealed that there is a pressure effect on this
important property also.\cite{khasanov} It was shown that the
in-plane penetration depth data of YBa$_{2}$Cu$_{2}$O$_{8}$ is
consistent with 3D-XY critical behavior,\cite{khasanov} where the
penetration depth and correlation length in direction $i$ diverge
near $T_{c}$ as $\lambda _{i}^{2}\left( T\right) =\lambda
_{i0}^{2}t^{-\nu }$ and $\xi _{i}\left( T\right) =\xi _{i0}t^{-\nu
}$ with $t=1-T/T_{c}$ and $\nu \simeq 2/3$. Moreover, as listed in
Table I, the critical amplitude $\lambda _{ab0}^{-2}$ was found to
be pressure dependent. The consistency with 3D-XY critical
behavior implies that the transition temperature $T_{c}$ and the
critical amplitudes of the in-plane penetration depths $\lambda
_{ab0}$ and the $c$-axis correlation lengths $\xi _{c0}$ are not
independent but related
by\cite{tshk,tsjh,book,parks,tspstat,peliasetto}

\begin{equation}
k_{B}T_{c}=\frac{\Phi _{0}^{2}}{16\pi ^{3}}\frac{f\xi
_{c0}}{\lambda _{ab0}^{2}},\text{ }f\simeq 0.453.  \label{eq8}
\end{equation}
Universality implies that this relation holds irrespective of the
applied pressure. Thus, in addition to Eq.(\ref{eq6}) the pressure
induced relative changes are then related by
\begin{equation}
\frac{\Delta T_{c}}{T_{c}}=-2\frac{\Delta \lambda _{ab0}}{\lambda
_{ab0}}+\frac{\Delta \xi _{c0}}{\xi _{c0}}  \label{eq9}
\end{equation}
and, because of $\gamma =\xi _{ab0}/\xi _{c0}$,
\begin{equation}
\frac{\Delta \gamma }{\gamma }=\frac{\Delta \xi _{ab0}}{\xi
_{ab0}}-\frac{\Delta \xi _{c0}}{\xi _{c0}}.
 \label{eq10}
\end{equation}
Using then the estimates for $\Delta T_{c}/T_{c}$ and $-2\Delta
\lambda _{ab0}/\lambda _{ab0}$ the relative change $\Delta \xi
_{c0}/\xi _{c0}$  is readily calculated with aid of
Eq.~(\ref{eq9}), while the values for $\Delta \xi _{ab0}/\xi
_{ab0}$ follow from relation (\ref{eq10}). The values of $\Delta
\xi _{c0}/\xi _{c0}$ and $\Delta \xi _{ab0}/\xi _{ab0}$  are
listed in Table \ref{Table:pressure_results}.

From the scaling analysis of the pressure effect on magnetization
and in-plane penetration depth it then emerges that the rise of
the transition temperature of underdoped YBa$_{2}$Cu$_{4}$O$_{8}$
with increasing pressure is associated with a decreasing
anisotropy and volume $V$. The relative change of the transition
temperature $\Delta T_{c}/T_{c}$ is seen to mirror essentially
that of the anisotropy $\Delta \gamma /\gamma $. This is
consistent with the generic behavior for high-temperature
superconductors, where for a given HTS family $T_{c}$ increases
with reduced aniostropy.\cite{parks,tshknj,tspstat} Although these
changes are small compared to those in the critical amplitudes of
the in-plane penetration depth $\Delta \lambda _{abo}/\lambda
_{abo}$ and the correlation lengths $\Delta \xi _{ab0}/\xi
_{ab0}$, $\Delta \xi _{c0}/\xi _{c0}$ (see Table
~\ref{Table:pressure_results}) it becomes clear that the pressure
induced change of the anisotropy is the essential ingredient which
remains to be understood microscopically. Empirically the
anisotropy decreases rather steeply by approaching optimum doping
and levels off in the overdoped regime.\cite{parks,tshknj,tspstat}
Together with Eq.(\ref{eq6}) this explains why the pressure effect
on $T_{c}$ becomes very small in optimally and overdoped cuprate
superconductors.\cite {takahashi,wijngaarden,klehe}

This work was partially supported by the Swiss National Science
Foundation, the NCCR program {\it Materials with Novel Electronic
Properties} (MaNEP) sponsored by the Swiss National Science
Foundation.

\end{document}